\documentclass[prb,aps,floats,twocolumn]{revtex4}

\usepackage{graphicx}
\usepackage{amsfonts}      % Paquete AMS
\usepackage{amssymb}      % Paquete AMS
\usepackage{amsbsy}      % Paquete AMS
\usepackage{amsmath}
\usepackage{xcolor}
\usepackage{hyperref}
\usepackage{titlesec}
\usepackage{bm}
\usepackage{ulem}

\makeatletter

\renewcommand{\Im}[1]{{\mathrm{Im}\,{#1}}}

\newcommand{\pair}[1]{\langle{#1}\rangle}

\newcommand{\normal}[1]{\,:\!{#1}\!:\,}

\DeclareMathOperator{\sgn}{sgn}
\DeclareMathOperator{\tr}{Tr}

\makeatother

\begin{document}
\title{Large-$S$ limit of the large-$N$ theory for the triangular antiferromagnet}

\author{Shang-Shun Zhang$^{1}$, E. A. Ghioldi$^{2}$, Yoshitomo Kamiya$^{3}$, L. O. Manuel$^{2}$, A. E. Trumper$^{2}$  and C. D. Batista$^{1,4}$}

\affiliation{$^1$ Department of Physics and Astronomy, University of Tennessee, Knoxville, Tennessee 37996-1200, USA}
\affiliation{$^2$ Instituto de F{\'i}sica Rosario (CONICET) and Universidad Nacional de Rosario, Boulevard 27 de Febrero 210 bis, (2000) Rosario, Argentina}
\affiliation{$^3$ School of Physics and Astronomy, Shanghai Jiao Tong University, Shanghai 200240, China,}
\affiliation{$^4$ Neutron Scattering Division and Shull-Wollan Center, Oak Ridge National Laboratory, Oak Ridge, Tennessee 37831, USA}

\begin{abstract}
Large-$S$ and large-$N$ theories (spin value $S$ and spinor component number $N$) are complementary, and sometimes conflicting, approaches to quantum magnetism.  While large-$S$ spin-wave theory captures the correct semiclassical behavior, large-$N$ theories, on the other hand, emphasize the quantumness of spin fluctuations. In order to evaluate the possibility of the non-trivial recovery of the semiclassical magnetic excitations within a large-$N$ approach,  we compute the large-$S$ limit of the dynamic spin structure of the triangular lattice Heisenberg antiferromagnet within a Schwinger boson spin representation.  We demonstrate that, only after the incorporation of Gaussian ($1/N$) corrections to the saddle-point ($N=\infty$) approximation, we are able to exactly reproduce the linear spin wave theory results in the large-$S$ limit. 
The key observation is that the effect of $1/N$ corrections is to cancel out exactly the main contribution of the saddle-point solution;  while the collective modes  (magnons) consist of two spinon bound states arising from the poles of the RPA propagator.  This result implies that it is essential to consider the interaction of the spinons with the emergent gauge fields and that the magnon dispersion relation should not be identified with that of  the saddle-point spinons.\\ 

\end{abstract}

\maketitle

\section{Introduction}  

Understanding the role of quantum fluctuations in frustrated antiferromagnets  has been the focus of multiple studies over the last decades.~\cite{Misguich2005,Sachdev2008,Normand2009,Balents2010,Powell2011,Lacroix2010,Savary2017,Zhou2017} These efforts were originally  motivated  by  the  resonant valence bond (RVB) state proposed by P. W. Anderson for describing the ground state of the triangular antiferromagnetic (AF) Heisenberg model.\cite{Anderson1973,Fazekas1974} The RVB state is a linear superposition of different configurations of short range singlet pairs, a quantum spin liquid state, whose resonant character leads to the decay of spin-$1$ 
excitations into pairs of free spin-$1/2$ spinons. This strongly quantum mechanical scenario has no classical counterpart, given that semi-classical phases correspond to magnetically ordered states with integer spin-$1$ excitations known as magnons.\cite{Mattis1981}
 
While the semiclassical picture relies on the spin wave theory\cite{Mattis1981,Auerbach1994} (large-$S$ expansion), a systematic and controlled  approach to the RVB picture can be formulated in the context of large-$N$ theories. Here the $SU(2)$ Heisenberg model is extended to a family of $SU(N)$ models, with $N$ being the number of  {\it flavors} of a  generalized spinor. In this formulation, the spin degree of freedom is represented by a product of spin-$\frac{1}{2}$ parton operators with bosonic (Schwinger) or fermionic (Abrikosov) character, subject to certain constraints.\cite{Baskaran87,Baskaran88,Arovas1988,Affleck1988,Read1991,Sachdev1991,Auerbach1994,Timm1998,Flint2009} The resulting Hamiltonian is expressed in terms of {\it isotropic} bond operators that emphasize the quantum nature of the bonds. The basic strategy is to describe the low-energy properties of the system, such as the dynamical spin susceptibility, by expanding the parameter $1/N$. The first term of the expansion corresponds to the saddle point (SP)  approximation, which is equivalent to the mean field theory, consisting of a gas of free spin-$\frac{1}{2}$ spinons. The  $1/N$ corrections introduce interactions between spinons mediated by emergent gauge fields.\cite{Arovas1988,Read1991,Sachdev1991,Auerbach1994,Chubukov1995} In the extreme $N\to \infty$ limit, the physics of  free spin-$\frac{1}{2}$ spinons associated to the SP solution is exact; while  the inclusion of $1/N$ corrections may drastically change  the SP physics  for finite $N$. 

Although large-$N$ treatments were introduced to describe quantum spin liquid states,\cite{Read1991,Sachdev1991,Affleck1988} there is a renewed interest focused on the reliability of the parton method for describing the excitation spectrum of magnetically ordered states near a quantum melting point (QMP). This is mainly motivated by the increasing number of magnetically ordered quantum magnets, whose excitation spectrum is not well described by a simple large-$S$ expansion.~\cite{Coldea2001,Coldea2003,Isakov05,ito2017structure,Ma2016,kamiya2018nature} In this context, the large-$N$ theory based on the Schwinger bosons (SB) representation is more adequate since, unlike the fermionic case, it can describe the magnetically ordered states through the condensation of the SBs.\cite{Hirsch1989,Sarker1989,Chandra1990} At the SP level, which is equivalent to the the Schwinger boson mean field theory (SBMFT), the dynamical spin susceptibility shows a two free-spinon continuum (branch cut) which misses the true collective modes (magnon) of the magnetically ordered state.\cite{Auerbach1988,Auerbach1994}   The main signal of the magnetic spectrum   is a pole located at the lower edge of the two-spinon continuum, that has the single-spinon dispersion. For collinear antiferromagnets and for a particular mean field decoupling of the Heisenberg term, this single-spinon dispersion  coincides with the semiclassical linear spin wave result. This coincidence was originally interpreted as a general attribute of the SBMFT.~\cite{Auerbach1988} However, it was later recognized that the single-spinon band (low energy edge of the  continuum) predicted by the SBMFT for {\it non-collinear} phases does not coincide with the single-magnon dispersion in the large-$S$ limit.  This fact was interpreted as a strong failure of the SBMFT.~\cite{Chandra1991,Coleman1994}  

As we will see later, this problem has a common root with the $O(4)$ symmetry of the fixed point at the transition between the non-collinear magnetic ordering and a gapped Z$_2$ spin liquid phase.~\cite{Chubukov1994S} The bottom line is that the gapless spinon modes of the SP solution have {\it the same spin velocity}.
This property, which is a direct consequence of the invariance of the Hamiltonian under global spin rotations and under spatial inversion, has two important implications. The first implication is that the long wavelength limit of  the SP action has an emergent $O(2N)$ symmetry. This symmetry is broken by the interaction terms (higher order terms in a $1/N$ expansion) that become irrelevant at the quantum critical point that signals the onset of the Z$_2$ quantum spin liquid phase.~\cite{Chubukov1994S}  The second implication is that the three gapless magnetic modes that are obtained from SP solution must also have the same velocity. This degenerate triplet of Goldstone modes is not consistent with the generic low-energy spectrum of  non-collinear magnets because the three Goldstone modes cannot be related by symmetry transformations. In particular, for the 120 degree ordering of the triangular Heisenberg antiferromagnet,  the velocity of the Goldstone mode associated with rotations about the axis perpendicular to the plane of the magnetic ordering is different from the velocity of the other two Goldstone modes.~\cite{Dombre89}

Motivated by these observations,  here we demonstrate  that  the LSWT result for the dynamical spin susceptibility is recovered in large-$S$ limit  upon adding a $1/N$ correction to the SP or SBMFT. 
For simplicity, we focus on the triangular lattice Heisenberg antiferromagnet with a $120^{\circ}$ N\'eel ground state ordering,  whose quantum ($S=1/2$) magnetic excitation spectrum is very different from the semiclassical ($S \to \infty$ ) limit.\cite{zheng2006excitation,Chernyshev2009,Mourigal2013a}

We have recently computed the dynamical spin structure factor of the $S=1/2$ triangular lattice antiferromagnet by  including $1/N$ corrections (Gaussian fluctuations) around the SP  solution.~\cite{Ghioldi2018} The predicted excitation spectrum reveals a strong quantum character consistent with a magnetically ordered ground state in the proximity of a QMP. The low energy part of the spectrum consists of two-spinon bound states (magnons) induced by fluctuations of the gauge fields, that emerge as poles of the RPA propagator.
A crucial observation is that the main signal of the SP solution (pole at the lower edge of the two-spinon continuum) is exactly canceled by the $1/N$ correction and the remaining low-energy poles are the poles of the RPA propagator. In view of this result, it is not surprising that the poles of the SBMFT theory do not coincide with the poles of the linear spin wave theory (LSWT) in the large-$S$ limit.~\cite{Chernyshev2009,Zhitomirsky2013}
In other words, magnons (collective modes of the underlying magnetically ordered ground state) should not be identified with the poles that appear in the dynamical spin susceptibility {\it at the SP level} (lower edge of two-spinon continuum),  but with the new poles  (poles of the RPA propagator) that appear in the dynamical spin susceptibility upon adding higher order $1/N$ corrections.  In Ref.~\onlinecite{Ghioldi2018} we demonstrated that, even for $S=1/2$ (quantum limit), the spin velocities of these poles basically coincide with the spin-wave velocities obtained from LSWT plus $1/S$ corrections.~\cite{Chubukov1994S,Chernyshev2009,Zhitomirsky2013} In this work we demonstrate these poles coincide over the full Brillouin zone with the ones obtained from LSWT  in the $S \to \infty$ limit. Furthemore, the spectral weight of the magnon peaks predicted by LSWT is also exactly recovered by the SBMFT plus a $1/N$ correction.

The article is organized as follows: Sec.~\ref{Large-N} is a general introduction to the large-$N$ Schwinger boson theory for frustrated antiferromagnets. More specifically, we review the extension  to $N>2$ that was proposed by Flint and Coleman,~\cite{Flint2009} by requiring that the generalized spin operators must preserve their transformation properties under 
rotations and under the time reversal operation. Sec.~\ref{SP}  describes the large-$N$ expansion of the extended theory around the SP solution. In Sec.~\ref{DSS} we present a formal $1/N$ expansion of the dynamical spin susceptibility. In particular, we discuss the four different Feynman diagrams that appear to order $1/N$.  In Sec.~\ref{SU2} we fix $N=2$ to consider the excitation spectrum of 
triangular lattice Heisenberg antiferromagnetic model, whose ground state is known to exhibit  $120^{\circ}$ N\'eel order
and take the large-$S$ limit (for fixed $N$) of the SP solution and the higher order $1/N$ corrections.
The results of Sec.~\ref{SU2} are applied in Sec.~\ref{DSS_largeS} to demonstrate that the dynamical spin structure factor predicted by  LSWT is exactly recovered when we add a particular $1/N$ correction (one of the four Feynman diagrams of Fig. \ref{fig2:feymann}) to the SP result. This is the $1/N$ correction that was recently included in Ref.~[\onlinecite{Ghioldi2018}].
We conclude the work in Sec.~\ref{Disc} with a general discussion of the implications of our result for other frustrated magnets.

\section{Large-N Schwinger boson theory for frustrated antiferromagnets
\label{Large-N}}

In this section we describe  the extension of the Schwinger boson theory for frustrated antiferromagnets  to arbitrary number of flavors $N$. For this purpose, we use the time reversal (symplectic) scheme introduced in Ref.~[\onlinecite{Flint2009}].  This discussion  complements the results presented  in Ref.~[\onlinecite{Ghioldi2018}]  for $N=2$ and $S=1/2$, by taking explicitly into account the $N$ dependence in the theory.  We start  by considering the  Schwinger boson representation of the   generators of $SU(N)$: $S_{\alpha\beta}=b^{\dagger}_{\alpha}b_{\beta}$  with $\alpha$ running over $N$ different flavors.~\cite{Auerbach1994} Following Ref.~[\onlinecite{Flint2009}], we will request that the large-$N$ theory must preserve not only the invariance of the Hamiltonian under time reversal and spin rotations, but also the properties of the generalized spins under these transformations.
The generators of $SU(N)$ can be divided into even and odd under a time reversal transformation. The odd ones are the generators of the $Sp(N)$ subgroup of $SU(N)$.
In the physical case $N=2$, the   isomorphism between $SU(2)$  and the simplectic $Sp(2)$ group implies that the three generators of $SU(2)$ must be odd under time reversal. The situation is different for $N > 2$ because the number of generators of $Sp(N)$ is smaller than the number of generators of $SU(N)$. The generators of $Sp(N)$ can be constructed by taking the antisymmetric combination between a generator $S_{\alpha \beta}$ of $SU(N)$ and its time reversed counterpart $ \sgn{\alpha} \sgn{\beta}  S_{-\beta-\alpha}$ version,
\begin{equation}
 \mathcal{S}_{\alpha\beta}= b^{\dagger}_{\alpha}b_{\beta}-\sgn{\alpha} \sgn{\beta} \; b^{\dagger}_{-\beta}b_{-\alpha},
 \label{symplectic}
\end{equation}
where $N$ is assumed to be even  and   $\alpha  , \beta= \text{-}\frac{N}{2},...,\frac{N}{2}$.  Note that the number of independent generators of $Sp(N)$ is $N(N+1)/2$ because $\mathcal{S}_{\alpha\beta} = -\sgn{\alpha} \sgn{\beta} \; \mathcal{S}_{{-\alpha} {- \beta}}$. As shown in Ref.~[\onlinecite{Flint2009}],  the Heisenberg  Hamiltonian of the generalized symplectic spins turns out to be
\begin{eqnarray}
{\cal H} =  \sum_{\pair{ij}} \frac{2J_{ij}}{N}  \hat{\mathcal{S} }_{i} \cdot \hat{\mathcal{ S}}_{j}, 
% \label{Heis}
\end{eqnarray}
 where $J_{ij}$ is rescaled by $N$ to make $\cal{H}$ extensive in  $N$ and the generalized Heisenberg interaction is given by
\begin{equation}
\hat{\mathcal{S} }_{i} \cdot \hat{\mathcal{ S}}_{j}  \equiv 
\frac{1}{8} \sum_{\alpha \beta} \mathcal{S}_{\alpha\beta}(i)  \mathcal{S}_{\beta \alpha}(j) 
= \normal{{B}_{ij}^{\dagger} {B}_{ij}} - {A}_{ij}^{\dagger} {A}_{ij},
\label{heis}
\end{equation}
where
\begin{equation}
A^{\dagger}_{ij}  = \frac{1}{2} \sum_{\alpha} \sgn{\alpha} \;b^{\dagger}_{i\alpha} b^{\dagger}_{j  \bar{\alpha}},
 \;\;\;\;
B^{\dagger}_{ij}  = \frac{1}{2} \sum_{\alpha}  b^{\dagger}_{i\alpha} b_{j\alpha },
\label{bop}
\end{equation}
are $Sp(N)$ invariant bond operators $ (\bar{\alpha} \equiv - \alpha)$. The bond operators  $A^{\dagger}_{ij}$ create  $Sp(N)$ singlets, while the
$B^{\dagger}_{ij}$ operators  make them resonate. Furthermore, the Casimir operator of the symplectic spins is\cite{Flint2009}  \\
\begin{equation}
 \hat{\mathcal{ S}}^2_i= \frac{1}{4} n_{bi}(n_{bi}+N),
\end{equation}
with $n_{bi}=\sum_{\alpha} b^{\dagger}_{i\alpha} b_{i\alpha}$. The Casimir operator results from fixing $n_{bi}=NS$:

\begin{equation}
 \hat{\mathcal{S}}^2_i= \frac{1}{4} N^2 S(S+1).
\end{equation}
We note  that  Eq.~\eqref{heis} coincides with the two singlet bond structure of the $SU(2)$ Schwinger boson theory for N$=\!\!2$.~\cite{Ceccatto1993} In particular,  the condition of one  $SU(2)$ Schwinger boson per site, $n_{bi}=2S=1$, that corresponds to $S=1/2$, is recovered through the Casimir operator for $N=2$. This two singlet bond structure is  adequate to describe noncollinear magnetic orderings\cite{Manuel1998,Manuel1999} and to classify quantum spin liquid states with the projective symmetry groups.~\cite{Wang2006,Messio2013}\\

\section{Saddle point expansion}
\label{SP}
The partition function of the interacting symplectic spins can be expressed as a functional integral over coherent states,~\cite{Auerbach1994,Ghioldi2018}
\begin{multline}
 \mathcal{Z} = \int D [\overline b,b]D[\lambda] \ e^{\! -\int_{0}^{\beta} d\tau \! \left[ \sum\limits_{i \alpha} \overline b_{i \alpha}^{\tau} \partial_{\tau} b_{i \alpha}^{\tau} + \ \mathcal{H}(\overline b,b) \   \right] } \\
 \times e^{ -\int_{0}^{\beta} d\tau \  i \sum\limits_{i} \lambda_{i}^{\tau} \big(\sum\limits_{\alpha} \overline b_{i \alpha}^{\tau} b_{i \alpha}^{\tau} - NS \big)  },
\label{partition}
\end{multline}
with a generalized spin Hamiltonian 
\begin{equation}
\mathcal{H}= \sum_{\langle i j \rangle} \frac{2 J_{ij}}{N} (\overline B_{ij}^{\tau } B_{ij}^{\tau} - \overline A_{ij}^{\tau} A_{ij}^{\tau} ).
\end{equation}
The integration measures are $ D[\overline b,b] = \prod_{i \tau \alpha} \frac{d\bar b _{i\alpha}^{\tau} db_{i\alpha}^{\tau}}{2 \pi i}$,  and $ D[\lambda] = \prod_{i \tau} \frac{d\lambda_{i}^{\tau}}{2 \pi}$. The local constraint, $n_{bi}=NS$, is incorporated via integration over the time- ($\tau$) and space- ($i$) dependent auxiliary field $\lambda_{i}^{\tau}$.\\

After introducing the Hubbard-Stratonovich (HS) transformations that decouple the $\overline{A} A$ and $\overline{B}B$ terms,~\cite{Ghioldi2018} the partition function becomes
\begin{equation}
 \mathcal{Z} = \int D[\overline W,W]D[\lambda] \ e^{-NS_{\rm eff}(\overline W, W, \lambda)},
\label{Seff}
\end{equation}
where the parameter $1/N$ plays the  role of the Planck's constant in a semiclassical expansion.
 $W\!\!=\!W^A,W^B$ are the space and time-dependent bond HS fields and the effective action is
\begin{eqnarray}
 S_{\rm eff}(\overline W, W, \lambda) =\! & \int_{0}^{\beta} & \! \!d\tau \sum\limits_{ijr} \frac{1}{2J_{ij}} \overline W_{ij}^{r\tau} W_{ij}^{r\tau}\!-\!iS \sum_{i} \lambda_{i}^{\tau} \nonumber \\
 &+& \frac{1}{N}  \tr \ln \left[ \mathcal{G}^{-1}(\overline W, W, \lambda) \right].
\label{effective}
\end{eqnarray}
\noindent The integration measure of the HS fields is $D[\overline W^{}\!,\!W] =  \prod_{ij\tau r} \frac{ d\overline {W}_{ij}^{r\tau} d W_{ij}^{r\tau}}{ 4 \pi i J_{ij}/N} $, with $r\!\!=\!\!A,B$, and  
$\mathcal{G}^{-1}\!\!\!\equiv\!\!\mathcal{M}$ is the bosonic dynamical matrix\cite{Ghioldi2018} with the trace taken over space, time, and  boson flavor indices. Note that the integration measure dependence on $J_{ij}$ has changed with respect to Ref.~[\onlinecite{Ghioldi2018}] in order to keep the factor of $N$ in front of $S_{\rm eff}$ [see Eq.\eqref{Seff}].

The effective action \eqref{effective} is invariant under a $U(1)$ gauge transformation of the SBs and the auxiliary fields.   The phase  of the HS fields $\overline{W}, W,$ and the Lagrange multiplier $\lambda$ represent the emergent gauge fields of the SB theory.~\cite{Auerbach1994}\\ 

To compute the partition function \eqref{Seff} we expand the effective action $S_{\rm eff}$ about its SP solution 
\begin{equation} \label{spcond}
S_{\rm eff} \equiv \sum_{n=0}^{\infty} \sum_{\alpha_1...\alpha_n}  S_{\alpha_1 \cdots \alpha_n}^{(n)}  \Delta \phi_{\alpha_1} \cdots \Delta \phi_{\alpha_n},
\end{equation} 
with
\begin{equation}\label{derivada}
S_{\alpha_1 \cdots \alpha_n}^{(n)}= \frac{1}{n!}\left.\frac{\partial^{n} S_{\rm eff}}{\partial \phi_{\alpha_1} \cdots \ \partial \phi_{\alpha_{n}}}\right|_{\rm sp},
\end{equation}
and $\Delta \phi_{\alpha}^{} = \phi_{\alpha} - \phi_{\alpha}^{\rm sp}$. The fields $\phi_{\alpha}$ are the auxiliary fields $\left\{ \overline W_{ij}^{r    \tau}, W_{ij}^{r \tau}, \lambda_{i}^{\tau} \right\}$ 
($\alpha$ includes field, space $i$, and time $\tau$ indices) and $\phi^{\rm sp}_{\alpha}$ is the SP solution 
 that fulfills the condition $S^{(1)}_{\alpha}=0$:
\begin{equation} \label{spcond}
\frac{\partial S_{\rm eff}}{\partial \phi_{\alpha}} \bigg|_{\rm sp} = 
\frac{\partial \  S_0}{\partial \phi_{\alpha}}\bigg|_{\rm sp} + \frac{1}{ N} \tr \bigg[\mathcal{G}_{\rm sp} \;v_{\alpha}\bigg] = 0,
\end{equation}
where $S_0$ is equal to the first  line of Eq.~\eqref{effective} and  $\mathcal{G}_{\rm sp}$ is the saddle point  Green's function and $v_{\alpha}=\frac{\partial \mathcal{G}^{-1}}{\partial \phi_{\alpha}}$ is the so-called internal vertex. $S^{(0)}$ coincides with the effective action $S^{\rm sp}_{\rm eff}$ evaluated at the SP solution, so the effective action can be rewritten as~\cite{Auerbach1994,Ghioldi2018}  
\begin{equation}\label{Seffexp}
 S_{\rm eff} = S^{\rm sp}_{\rm eff} + \sum_{\alpha_1 \alpha_2} S_{\alpha_1 \alpha_2}^{(2)} \ \Delta \phi_{\alpha_1} \Delta \phi_{\alpha_2}  + S_{\rm int},
\end{equation}
with
\begin{equation}
 S_{\rm int}=\sum_{n=3}^{\infty} \sum_{\alpha_1 \cdots \alpha_n} S_{\alpha_1 \cdots \alpha_n}^{(n)} \ \Delta \phi_{\alpha_1} \cdots \Delta \phi_{\alpha_n}\label{Sint}.
\end{equation}
It is straightforward to show that 
\begin{eqnarray}
 S^{\rm sp}_{\rm eff} =  &\int_{0}^{\beta}& \! d\tau \; ( \sum\limits_{ijr} \frac{1}{2J_{ij}} \overline W_{ij}^{r\tau} W_{ij}^{r\tau}\!-\!iS \sum_{i} \lambda_{i}^{\tau} ) \Big|_{\rm sp} \nonumber \\
 &+&\!\!\!\frac{1}{N}  \tr \ln \left[ \mathcal{G}_{\rm sp}^{-1} \right]\label{s0},
\end{eqnarray}
\begin{eqnarray}
 S^{(2)}_{\alpha \alpha{\prime}} &=&  \frac{1}{4J_{ij}}  (\delta_{\alpha,W^{r \tau}_{ij}} \delta_{\alpha^{\prime},\overline{W}^{r \tau}_{ij}} +\delta_{\alpha,\overline{W}^{r \tau}_{ij}} \delta_{\alpha^{\prime},{W}^{r \tau}_{ij}})  \nonumber \\ 
 & & - \frac{1}{2N}  \tr \ln \left[\; \mathcal{G}_{sp}\; v_{\alpha 
 } \; \mathcal{G}_{sp} \; v_{\alpha^{\prime} } \right]\label{s2} ,
\end{eqnarray}
and

\begin{equation}
 S^{(n\ge 3)}_{\alpha_1... \alpha_{n}} = \frac{(\text{-}1)^{n+1}}{n!\; n} \!\! \sum_{P(\alpha_1...\alpha_n)}\!\! \frac{1}{N} \tr \ln \left[\; \mathcal{G}_{\rm sp}\; v_{P_1 
 }\;...\; \mathcal{G}_{\rm{sp}} \; v_{P_{n} } \right]\label{sn} ,
\end{equation}
where $P(\alpha_1...\alpha_n)$ denotes all the different permutations of $(\alpha_1...\alpha_n)$. $S^{\rm sp}_{eff}$, $S^{(2)}_{\alpha \alpha{\prime}}$, and $S^{(n\ge 3)}_{\alpha_1... \alpha_{n}}$ are all of order $N^0$  because the trace
over the flavor index that appears in equations (\ref{s0})-(\ref{sn})  scales like $N$. \\

$S_{\rm int}$ is neglected from Eq.~\eqref{Seffexp}   at the Gaussian level and the free energy $F=-\frac{1}{\beta} \ln \mathcal{Z}$ per flavor becomes
\begin{equation}
 \frac{F^{(2)}}{N}=\frac{1}{\beta}S^{sp}_{\rm eff} -\frac{1}{N\beta} \tr \ln \left[  S^{(2)} \right],
\label{free}
 \end{equation}
with $\beta=1/T$. 
Here the trace must be computed over time, space, and the auxiliary field index. Consequently, the contribution of the Gaussian fluctuations to the free energy per flavor is of order $1/N$.

\section{Dynamical spin susceptibility: 1/N expansion}
\label{DSS}

The computation of the dynamical spin susceptibility requires to couple the symplectic spins (\ref{symplectic}) to  space  and time-dependent external sources  $j^{\tau }_{i \alpha \beta}$
\begin{equation}
 \mathcal{J}_s= \sum _{i} j^{\tau }_{i \alpha \beta} \mathcal{S}^{\tau}_{\beta \alpha}(i),
 \label{source}
\end{equation}
where the sum over repeated flavor indices is assumed. After adding this term to the Lagrangian in Eq.~\eqref{partition}, the dynamical susceptibility is obtained from the generatriz $\mathcal{Z}[j]$\cite{Auerbach1994,Ghioldi2018}
\begin{equation}
 \chi_{\alpha\beta}(1,2) = \frac{\partial^{2} ln \mathcal{Z}[j]}{\partial j_{1}^{ \alpha \beta} \partial j_{2}^{\ \beta\alpha}} \Big|_{j=0} ,
\end{equation}
\begin{figure}[!t]
 \includegraphics[scale=0.5]{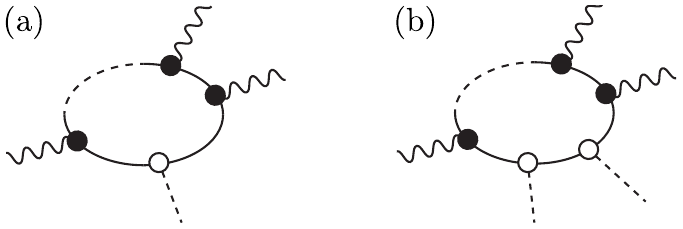}
 \caption{Diagrammatic representation of the external loops corresponding to one external vertex  $S^{(n+1)}$ (a), and two external vertices $S^{(n+2)}$ (b).~\cite{Auerbach1994}}
 \label{external-loops}
 \end{figure}
where $1$ and $2$ denote space and time points, $\bm r_1$,  and $\bm r_2$, respectively. The above expression can be split into two contributions,
\begin{equation}
 \chi = \chi_{_I} + \chi_{_{II}}, 
\end{equation}
with
\begin{eqnarray}\label{chi1}
 \chi_{_{I} \alpha\beta}(1,2)\!\! &=&\! \! \frac{N}{\mathcal{Z}}\int \!\!D[\overline \phi, \phi] \Big(\!-\!\frac{\partial^2S_{\rm eff}}{\partial j^{\alpha \beta}_{1}\partial j^{\beta \alpha}_{2}}  
 \Big|_{j=0}\; \Big)  \nonumber\\  &&\times  \ \ e^{-NS_{\rm eff}(\overline \phi, \phi,j=0)}   
\end{eqnarray}
and
\begin{eqnarray}\label{chi2}
 \chi_{_{II} \alpha\beta}(1,2)\! &=&\! \! \frac{N^2}{\mathcal{Z}}\!\!\int \!\!D[\overline \phi, \phi] \Big(\frac{\partial S_{\rm eff}}{\partial j^{\alpha \beta}_{1}} \Big|_{j=0} \frac{\partial S_{\rm eff}}{\partial j^{\beta \alpha}_{2}} \Big|_{j=0} \Big)  \nonumber\\  &&\times  \ \ e^{-NS_{\rm eff}(\overline \phi, \phi,j=0)}.   
\end{eqnarray}
The partial derivatives of the effective action are given by
\begin{eqnarray}
 N \frac{\partial S_{\rm eff}}{\partial j^{\alpha \beta }_{1}}\Big|_{j=0} &=& \tr\;[\;\mathcal{G}(j=0)\; u^{\alpha \beta}(1) \;],
\nonumber \\
 N \frac{\partial^2S_{\rm eff}}{\partial j^{\alpha \beta}_{1}\partial j^{\beta \alpha}_{2}}  
 \Big|_{j=0}\!\! &=&
 \tr [ \mathcal{G}(j\!=\!0)\; u^{\alpha \beta}(1)  \times  \mathcal{G}(j\!=\!0)\; u^{\beta \alpha}\!(2)\;],
 \nonumber \\
\end{eqnarray}
where $u^{\alpha \beta}(1)\equiv \partial \mathcal{G}^{-1}/\partial j^{\alpha \beta}_{1}$ is the so-called external vertex.
By using the SP expansion \eqref{Seffexp} and defining 
\begin{equation}
 S^{(n+1)}_{\alpha_1...\alpha_n;(\bm r_i ; \alpha \beta)}= N 
 \frac{\partial S^{(n)}_{\alpha_1...\alpha_n}(j)}{\partial j^{\alpha \beta}_{i}} \Big|_{j=0}
 \label{Sn1}
\end{equation}
and 
\begin{equation}
 S^{(n+2)}_{\alpha_1...\alpha_n;(\bm r_1\; \alpha \beta),(\bm r_2; \beta \alpha)}= N 
 \frac{\partial^2 S^{(n)}_{\alpha_1...\alpha_n}(j)}{\partial j^{\alpha \beta}
 _{1} \partial j^{\beta \alpha}_{2}} \Big|_{j=0},
 \label{Sn2}
\end{equation}
 which are diagrammatically represented in Fig.~\ref{external-loops}, we obtain an explicit expansion of $\chi_{_{I} \alpha\beta}(1,2)$ and $\chi_{_{II} \alpha\beta}(1,2)$ [Eqs. \eqref{chi1} and \eqref{chi2}] in powers of $1/N$:
\begin{widetext}
\begin{equation}
\chi_{_{I} \alpha\beta}(1,2)\!=\! \frac{1}{\mathcal{Z}}\! \int [D\overline \phi D\phi] \Big(\! \text{-} \sum^{\infty}_{n= 0} S^{(n+2)}_{\alpha_1...\alpha_n;(1 \; \alpha \beta),(2 \; \beta \alpha)} \Delta \phi_{\alpha_1}...\Delta \phi_{\alpha_n}\Big) \times \Big[ 
\sum^{\infty}_{L=0}\frac{(\text{-}N)^L}{L!} {(S_{\rm int})^L}
\Big]  e^{-N \left(\Delta \phi_{\alpha} S^{(2)}_{\alpha \alpha^{\prime}} \Delta \phi_{\alpha^{\prime}}+S^{sp}_{\rm eff} \right)}\label{chiI}
\end{equation}

\begin{eqnarray}
\chi_{_{II} \alpha\beta}(1,2)\!=\! \frac{1}{\mathcal{Z}}\! \int [D\overline \phi D\phi] \Big(\! \text{-} \sum^{\infty}_{ n= 0} S^{(n+1)}_{\alpha_1...\alpha_n;(1 \; \alpha \beta)} \Delta \phi_{\alpha_1}...\Delta \phi_{\alpha_n}\Big) 
&\times& \Big(\! \text{-} \sum^{\infty}_{n= 0} S^{(n+1)}_{\alpha_1...\alpha_m;(2 \; \beta \alpha)} \Delta \phi_{\alpha_1}...\Delta \phi_{\alpha_m}\Big) \nonumber \\ 
 &\times& \Big[ 
\sum^{\infty}_{L=0}\frac{(\text{-}N)^L}{L!}  {(S_{\rm int})^L}
\Big]  e^{-N \left(\Delta \phi_{\alpha} S^{(2)}_{\alpha \alpha^{\prime}} \Delta \phi_{\alpha^{\prime}}+ S^{sp}_{\rm eff}\right)} \label{chiII}
\end{eqnarray}
\end{widetext}
where 
\begin{equation}
\mathcal{Z}\!=\!\int [D\overline \phi D\phi] \Big[\!  
\sum^{\infty}_{L=0}\frac{(\text{-}N)^L}{L!}  {(S_{\rm int})^L}
\Big] \; e^{-N \left(\Delta \phi_{\alpha} S^{(2)}_{\alpha \alpha^{\prime}} \Delta \phi_{\alpha^{\prime}}+S^{sp}_{\rm eff} \right)}.
\end{equation}
Note that diagrams that arise from $\chi_{_{I} \alpha\beta}$ must contain two external lines that arrive to the same loop [see Fig.~\ref{external-loops}~(b)]. In contrast, the two external lines of the diagrams that arise from $\chi_{_{II} \alpha\beta}$ are connected to different loops.
The integrals of an even number of fields $\phi$ is the sum of all possible pair contractions (Wick's theorem) that defines the RPA propagator $ {D}_{\alpha_1\alpha_2}= [ 2 S^{(2)}]^{-1}_{\alpha_1\;\alpha_2}$:
\begin{equation}
\frac{1}{N}D_{\alpha_1\alpha_2}=
\frac{1}{\cal{Z}} \int [D\overline \phi D\phi]\;  \phi_{\alpha_1} \phi_{\alpha_2} \; e^{-N \Delta \phi_{\alpha} S^{(2)}_{\alpha \alpha^{\prime}} \Delta \phi_{\alpha^{\prime}}}.
\end{equation}
The diagrams for $\chi_{_I}$ and $\chi_{_{II}}$ [see Eqs. \eqref{chiI} and \eqref{chiII}] are constructed as follows\cite{Auerbach1994}:
the elements $S^{(n+1)}$ and $S^{(n+2)}$ contribute to external loops with $n$ internal vertices and one and two external vertices, respectively. The derivatives of $S^{(n)}$ [see Eqs.~\eqref{s0}-\eqref{sn}] with respect to  $j_i$ are of order $1/N$ because each source $j_i$ carries a flavor index.
Consequently, according to the  definition of $S^{(n+1)}$ and $S^{(n+2)}$ given by Eqs.~\eqref{Sn1} and \eqref{Sn2}, these external loops are of  order $N^0$.  The terms of the expansion of $S_{\rm int}$ in Eq.~\eqref{Sint} contribute to internal loops with $n\ge3$ internal vertices. Even though  $S_{\rm int}$ is of order $N^0$,  it is  multiplied by factor $N$, implying that each diagram contains a factor $N^L$, where $L$ is the number of internal loops. In addition, each contraction of the $\phi$ fields  gives rise to an RPA propagator $D$ (of order $N^0$) divided by $N$. Summarizing, each external loop contributes with a factor of order $N^0$, each internal loop contributes with a factor  of order $N$ and each  field contraction  contributes with a factor $1/N$. In other words, a diagram with $L$ internal loops and $P$ RPA propagators is of order $(\frac{1}{N})^{P-L}$.  Fig.~\ref{fig2:feymann}(a) shows the SP contribution to the dynamical susceptibility, while  Figs.~\ref{fig2:feymann}(b)-(e) show all the diagrams of order $1/N$  that contribute to  $\chi_{_{I}}$ and $\chi_{_{II}}$.
In particular, the diagram of Fig.~\ref{fig2:feymann}(b) corresponds to $\chi_{_{II}}$ for $L=0$ and $P=1$, while Figs.~\ref{fig2:feymann}(c) and (d) are the diagrams corresponding to $\chi_{_{I}}$ for $L=0$ and $P=1$. The diagram shown in Fig. ~\ref{fig2:feymann}(e)  corresponds to $\chi_{_{I}}$ and it is the only diagram that arises from non-Gaussian corrections of the effective action with  one internal loop ($L=1$) and two RPA propagators ($P=2$). 

\begin{figure}[!t]
 \includegraphics[width=\columnwidth,bb=0 0 481 318]{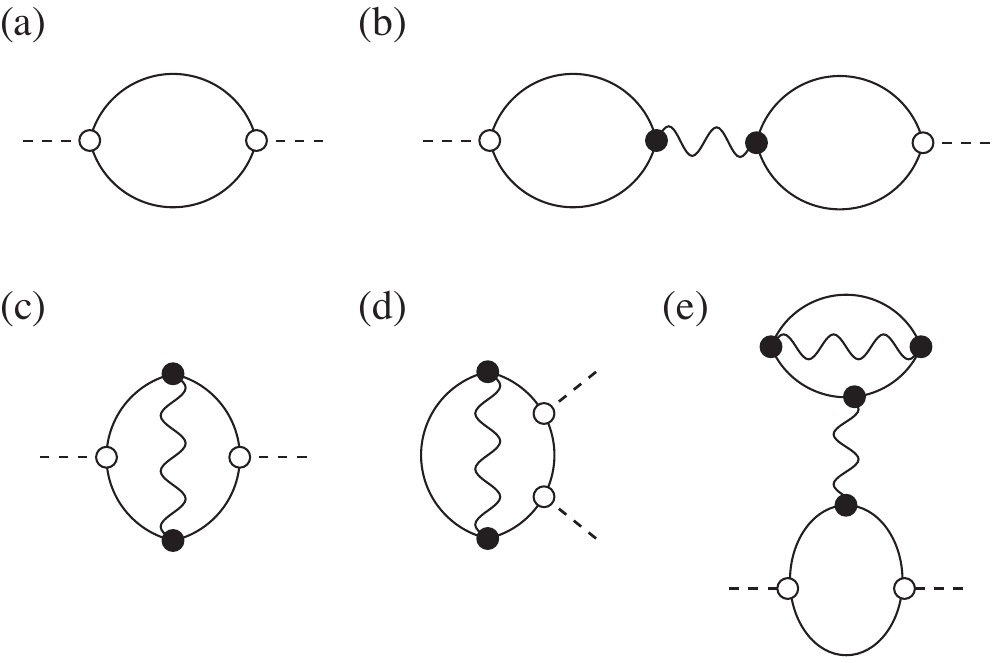}
 \caption{Diagrammatic representation of (a) saddle point contribution and  (b-e) the $1/N$ corrections to the dynamical spin susceptibility. 
 In our calculation we only include the contribution (b) for reasons explained in the text. The diagram (c) corresponds to a vertex correction
 relative to (a), while the diagrams (d) and (e) include a Hartree-Fock correction of the single-spinon propagator. The dashed lines represent the  external sources, the full lines represent  spinon propagators $\cal{G}^{\rm sp}$ at the SP level  and the wavy lines represent the RPA propagator $\frac{D}{N}$~\citep{Auerbach1994}. Solid and empty circles denote internal $u$ and external $v$ vertices, respectively.}
 \label{fig2:feymann}
 \end{figure}

\section{$SU(2)$ case: the $120^{\circ}$ N\'eel-ordered state}
\label{SU2}

The large-$N$ Schwinger boson theory developed in the previous sections is valid for the family of $Sp(N)$ models. Therefore, given that $SU(2)\! \cong \!Sp(2)$, the $SU(2)$ case is recovered by fixing $N\!\!=\!2$ in the above expressions.
 To study  the magnetic excitation spectrum of the 120$^{\circ}$ N\`eel-ordered ground state of the triangular SU(2) Heisenberg antiferromagnet,  we must add a symmetry breaking field, $h$, that selects the ordered ground state in the thermodynamic limit.~\cite{Ghioldi2018} The field $h$ couples linearly to  N\`eel order parameter and it is  set to zero after taking the thermodynamic limit. In the SB language, this process corresponds to condensing  the SBs in a single particle state (the single-spinon ground state is degenerate) that spontaneously breaks the $SU(2)$ symmetry of the spin Hamiltonian. 

 Only the diagram shown in Fig.~\ref{fig2:feymann}(a) contributes to the dynamical spin susceptibility at the SP level (SBMFT):
\begin{equation} 
\chi_{_I \mu\nu}^{\rm sp}(\bm q,i{\omega_n}) = \frac{1}{2}\tr
 \left[ \mathcal{G}^{\rm sp}  u^{\mu}(\bm q,i{\omega_n}) \mathcal{G}^{\rm sp} u^{\mu}(-\bm q,-i{\omega_n}) \right].\label{chispI}
\end{equation}
The index $\mu=x,y,z$ refers to the three spin components and $u^{\mu}$ is the external vertex that couples the spin excitations to the $\bm q$ component of an external magnetic field. It can be shown that $\chi_{_{II} \mu\mu}^{\rm sp}=0$.\cite{Ghioldi2018} 
 The magnetic excitation spectrum of $\chi_{_I \mu\mu}^{\rm sp}$ consists of a two-spinon continuum (branch cut), corresponding to a gas of free spin-$\frac{1}{2}$ spinons. The condensation of the SBs also generates a delta function contribution (pole) at the lower edge of the two-spinon continuum. In addition, due to the relaxation of the local constraint, the magnetic spectrum also exhibits  spurious modes arising from  density fluctuations of the SBs.\cite{Arovas1988,Auerbach1994,Mezio2011,Mezio2012,Ghioldi2018}  The inclusion of the $1/N$ correction  corresponding to the diagram shown in Fig.~\ref{fig2:feymann}(b) leads to the following contribution~\cite{Ghioldi2018} 
\begin{eqnarray}
\chi_{_{II}\; \mu\nu}^{\rm fl}(\bm q, i{\omega_n}) = \sum_{\alpha_1 \alpha_2}\!\!\!\! &&\frac{1}{2}\tr\big[ \mathcal{G}^{\rm sp} \ v_{\phi_{\alpha_1}} \ \mathcal{G}^{\rm sp} \ u^{\mu}(\bm q, i{\omega_n}) \big]  \nonumber \\
&&\;\;\;\;\;\;\; \times  \frac{1}{2}  D_{\alpha_{2} \alpha_{1}}(\bm q, i{\omega_n})  
\nonumber \\
&&\frac{1}{2} \tr\big[ \mathcal{G}^{\rm sp} \ v_{\phi_{\alpha_2}} \
\mathcal{G}^{\rm sp} \ u^{\nu}(-\bm q, -i{\omega_n}) \big]. 
\nonumber \\
\label{corr}
\end{eqnarray}
The factors of $1/2$ in front of each trace must be included to avoid double-counting of momenta (see below).
In  Ref.~[\onlinecite{Ghioldi2018}] we demonstrated that this particular $1/N$ correction introduces a drastic change in the dynamical spin susceptibility. In the first place, it cancels out the SP poles at the lower edge of the two-spinon continuum and it introduces new poles, which are the poles of the RPA propagator $D$. As we will show below, these new poles are associated with the collective modes (magnons) of the theory  and they correspond to two-spinon bound states generated by the fluctuations of the gauge fields. In the second place, the spurious modes of the SP solution are also exactly canceled out.
% {\color{blue}(the local constraint is imposed exactly to this $1/N$ order.}\cite{Raykin1993,Auerbach1994} 
%
It is important to note that the contribution from this diagram is exactly equal to zero for a singlet ground state ($h=0$).~\cite{Arovas1988} However, we have recently shown in Ref.~[\onlinecite{Ghioldi2018}] that it becomes finite for the magnetically ordered ground state under consideration. Moreover, for $N=2$ and $S=1/2$, the magnon dispersion obtained from this particular $1/N$ correction has Goldstone modes at the  $\Gamma$ and $\pm K$ points, whose velocities agree very well with the results obtained with LSWT plus $1/S$ corrections.~\cite{Chubukov1994S,Chernyshev2009} 

Below we demonstrate another virtue of this $1/N$ correction.   The LSWT result for the dynamical spin susceptibility, which  is exact in the large-$S$ limit, is recovered from the large-$N$ expansion by keeping only the diagrams shown in Figs.~\ref{fig2:feymann}~(a) and (b) [see Eqs.~(\ref{chispI}) and \eqref{corr}] and taking the $S \to \infty$ limit.

\subsection{Large-S limit}

The SP approximation is equivalent to the  SBMFT described by the quadratic  mean field Hamiltonian\cite{Mezio2011}
\begin{equation}\label{eq:sbmfH}
{\cal H}_{B} = \sum_k  \psi^{\dagger}_{\bm k}  {\cal H}_{MF} (\bm k)  \psi_k,
\end{equation}
 with $\psi_{\bm k} = (b_{{\bm k},\uparrow},b_{-{\bm k},\downarrow}^{\dagger})$,
\begin{equation}
 {\cal H}_{MF} ({\bm k})={
 \left(
\begin{array}{cc}
  \lambda_{sp} + \gamma_{\bm{k}}^{B}  &   -\gamma_{\bm{k}}^{A}   \\ \\
 -\gamma_{\bm{k}}^{A}  &   \lambda_{sp} + \gamma_{\bm{k}}^{B} 
\end{array} \right)} \ ,
\end{equation}
and
\begin{eqnarray}
 \gamma_{\bm k}^{A} &=& \sum\limits_{\bm \delta>0} J_{\delta} A_{\bm \delta} \sin \left(\bm{k\cdot \bm \delta}\right), \\   
 \gamma_{\bm k}^{B} &=& \sum\limits_{\bm \delta>0} J_{\delta} B_{\bm \delta} \cos \left(\bm{k\cdot \bm \delta}\right).
 \end{eqnarray}
The amplitudes $i A_{\bm \delta}$ and $B_{\bm \delta}$ are the SP values of the bond operators ${A}_{i,i+\bm \delta}$ and ${B}_{i,i+\bm \delta}$, while  $i\lambda_{sp}$ is the SP value of the Lagrange multiplier that was introduced to implement the local constraint 
${ b}^{\dagger}_{i\uparrow} {b}^{\;}_{i\uparrow} + {b}^{\dagger}_{i\downarrow} {b}^{\;}_{i\downarrow}=2 S$. 
The single-spinon Green's function is given by the $2$ by $2$ matrix
\begin{align}
G^{\rm sp}_0(\bm k, i \omega_n) =
\left(\begin{array}{cc}
{\lambda_{sp} + \gamma_{\bm k}^B +i\omega_n \over \varepsilon_{\bm k}^2 + \omega_n^2} & -{\gamma_{\bm k}^A \over \varepsilon_{\bm k}^2 + \omega_n^2}   \\
-{\gamma_{\bm k}^A \over \varepsilon_{\bm k}^2 + \omega_n^2} & {\lambda_{sp} + \gamma_{\bm k}^B -i\omega_n \over \varepsilon_{\bm k}^2 + \omega_n^2}  \\
\end{array}\right).
\end{align}
The poles of this Green's function determine the single-spinon dispersion, 
\begin{equation}
\varepsilon_{\bm k} = \sqrt{(\lambda_{sp} + \gamma_{\bm k}^B)^2 - (\gamma_{\bm k}^{A})^2}.
\end{equation}

\begin{figure}[t!]
\centering
\includegraphics[scale=0.4]{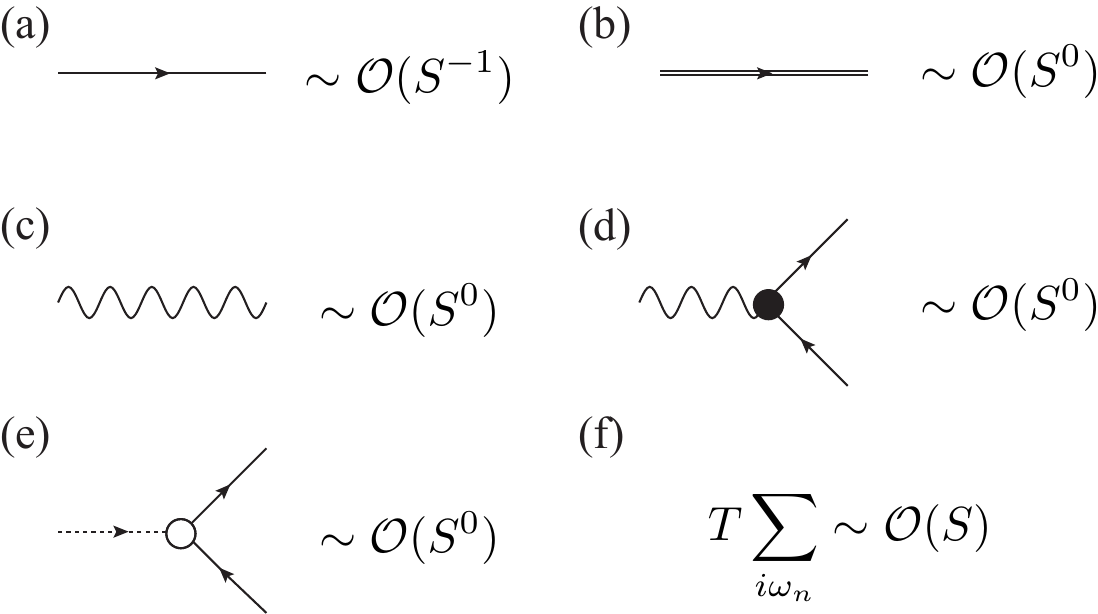}\caption{Power counting rule for the $S$ power of each Feynman diagram. Solid line:  non-condensed boson propagator ${\cal G}^{\rm sp}_0$.  Double line: condensed boson propagator ${\cal G}^{\rm sp}_c$.  Wavy line: RPA propagator ${D}$ of the fluctuation fields. Dashed line: external source. Solid and empty circles denote internal $v$ and external $u$ vertices, respectively.}
\label{powercounting}
\end{figure}

The SP single-spinon spectrum has two degenerate minima at ${\bm k} = \pm {{\bm Q} \over 2}$. On a finite size lattice, the minimum energy, $\varepsilon_{\pm {{\bm Q} \over 2}}$, is proportional to $1/{\cal N}_s$, where ${\cal N}_s$ is the number of lattice sites, and the ground state of ${\cal H}_B$ is a singlet state. Upon taking the thermodynamic limit, ${\cal N}_s \rightarrow \infty$, the spectrum becomes gapless at $\pm {{\bm Q} \over 2}$ and the bosons condense at $T=0$. Given that there are four single particle ground states (two gapless points with momenta ${\pm {{\bm Q} \over 2}}$ and two possible spin orientations), there is a continuous ground state degeneracy corresponding to different ways of condensing the  bosons.  The infinitesimal symmetry-breaking field $h$  selects a ground state with a particular $120^{\circ}$ magnetic ordering.~\cite{Ghioldi2018}
%The ground state of ${\cal H}_B$ is a singlet state on finite size lattices. However, the bosons condense in the thermodynamic limit leading to a continuous ground state degeneracy that reflects the spontaneous SU(2) symmetry breaking that leads to the emergence of magnetic ordering. 
It is then  convenient to work in the twisted spin reference frame where the selected 120$^{\circ}$ magnetic ordering becomes an in-plane ferromagnetic (FM) ordering along the $x$-axis. The real space Schwinger boson operators become $b_{i \uparrow} = \tilde{b}_{i \uparrow} e^{-i{\bm Q}\cdot {\bm r}/2}$ and $b_{i \downarrow} = \tilde{b}_{i \downarrow} e^{i{\bm Q}\cdot {\bm r}/2}$ in the new reference frame and the FM magnetic ordering arises from condensation  at momentum ${\bm k}={\bm 0}$. 
After taking the thermodynamic limit and sending $h$ to zero (the two operations do not commute), the  SP Green's  function of the spinons becomes
\begin{equation}
\mathcal{G}^{\rm sp} (\bm{k},i\omega_{n})=\mathcal{G}^{\rm sp}_{0}(\bm{k},i\omega_{n})+(2\pi)^{2}\delta(\bm{k}) \mathcal{G}^{\rm sp}_{c}(i\omega_{n}),\label{eq:GF_condensate}
\end{equation}
where $\mathcal{G}^{\rm sp}_{0}(\bm{k},i\omega_{n})$ and $\mathcal{G}^{\rm sp}_{c}(i\omega_{n})$ are the contributions from the non-condensed and  condensed bosons, respectively.  
After extending the two-component representation, $\psi_{\bm k}$, to the four-component representation $\Psi_{\bm k} = (b_{\bm k-{\bm Q}/2, \uparrow},\bar b_{ -\bm k + {\bm Q}/2 \downarrow}, b_{ \bm k + {\bm Q}/2 \downarrow}, \bar b_{-\bm k-{\bm Q}/2, \uparrow})$, we obtain
\begin{equation} \label{eq:green_nonc}
\mathcal{G}^{\rm sp}_{0}(\bm{k},i\omega_{n}) =
\left(  \begin{array}{cc}
 G^{\rm sp}_0(\bm k -{\bm Q \over 2},i\omega_n) & 0  \\
0 & G^{\rm sp}_0(-\bm k -{\bm Q \over 2},i\omega_n)
\end{array}\right),
\end{equation}
whose single-spinon pole locates at $\varepsilon_{\pm { \bm k} -{\bm Q \over 2}}$. For the condensed spinons, we have
\begin{equation} \label{eq:green_cond}
 \mathcal{G}^{\rm sp}_{c}(i\omega_{n})=
{n_c \Omega_c \over \Omega_c^2+\omega_n^2} \left(\!\!\begin{array}{cccc}
1 & 1 & -1   & -1  \\
1 & 1 & -1   & -1  \\
-1 &  -1 & 1 & 1 \\
-1 &  -1 & 1 & 1
\end{array}\!\!\right),
\end{equation}
where $\Omega_c = {h \over 2}$ and $n_c$ is the density of the condensate. 
The symmetry-breaking field is sent to zero  in the thermodynamic limit, meaning that $h = 0^+$.

\subsubsection{Large-$S$ limit of the saddle point solution}

 The self-consistent SP equations (\ref{spcond}), for arbitrary spin size $S$,  become:\cite{Ghioldi2018}

\begin{eqnarray}
&& B_{\bm{\delta}} = \int \frac{d^{2}\bm{k}}{(2\pi)^{2}}\cos(\bm{k}\cdot\bm{\delta}) {\lambda_{sp} + \gamma_{\bm k}^B \over 2\varepsilon_{\bm k}}+{n_c\over 2}\cos[\frac{\bm{Q}}{2}\cdot\bm{\delta}],\label{eq:sp1}
\nonumber \\
&& A_{\bm{\delta}}  = i\int\frac{d^{2}\bm{k}}{(2\pi)^{2}}\sin(\bm{k}\cdot\bm{\delta}) {\gamma_{\bm k}^A \over 2\varepsilon_{\bm k}}+i {n_c \over 2}\sin[\frac{\bm{Q}}{2}\cdot\bm{\delta}],
\label{eq:sp2}
\nonumber \\
&& 2S+1 = \int\frac{d^{2}\bm{k}}{(2\pi)^{2}} {\lambda_{sp} + \gamma_{\bm k}^B \over \varepsilon_{\bm k}} +n_c.
\label{eq:sp3}
\end{eqnarray}

\noindent  In all cases, the integral that appears in each of the three expressions is the contribution from the non-condensed spinons, while the second term, proportional to $n_c$, is the contribution from the condensate. 

In the large-$S$ limit, the ground state is the $120^{\circ}$ N{\'e}el-ordered state characterized by $\langle {\bm S}_i \rangle = S {\bm n}_i$ with ${\bm n}_i$ the unit vector along the local moments. In the SBMFT, $\langle {\bm S}_i \rangle = \frac{1}{2}\langle b_{i\alpha}\rangle^* {\boldsymbol \sigma}_{\alpha \beta} \langle b_{i \beta}\rangle $, where ${\boldsymbol \sigma}$ is the vector of Pauli matrices, implying that $\langle b_{i \beta}\rangle \sim \sqrt{S}$. This observation fixes  the scaling  of the SP parameters: $n_c = \langle b_{i\alpha}\rangle^*  \langle b_{i \alpha}\rangle \sim S$ and $A_{\bm \delta}, B_{\bm \delta} \sim S$, for large enough $S$, implying that $\gamma_{\bm k}^A$, $\gamma_{\bm k}^B$ and $\varepsilon_{\bm k}$  are also ${\cal O} (S)$.  Back to the saddle point equations \eqref{eq:sp3}, we observe that the contribution from the non-condensed bosons is of order $S^0$, while the contribution from the condensed bosons is of order $S$,   implying that these equations become much simpler,
\begin{align}
n_c \to 2S, \ \
B_{\bm{\delta}}  \to {S}\cos[\frac{\bm{Q}}{2}\cdot\bm{\delta}],  \ \  A_{\bm{\delta}}  \to i {S }\sin[\frac{\bm{Q}}{2}\cdot\bm{\delta}],
\label{eq:sp1}
\end{align}
\noindent in the $S\rightarrow \infty$ limit. Note that the saddle point value of the Lagrange multiplier is equal to $\lambda_{sp}={3\over 2}JS$, as required by the gapless nature of $\varepsilon_{\bm k}$.

 This solution indicates that the rescaled mean-field Hamiltonian $\tilde{{\cal H}}_B = {\cal H}_B S^{-1}$ and frequency $\tilde{\omega} = \omega S^{-1} $ are independent of $S$ or  ${\cal O}(S^0)$ in the large $S$ limit. Consequently,  each integral over frequency (summation over Matsubara frequencies) introduces an $S$ factor: $\int{d\omega \over 2\pi} = S \int{d\tilde{\omega} \over 2\pi}$. In addition, according to Eq.~(\ref{eq:green_nonc}),  the Green's function $\mathcal{G}_0$ of the non-condensed Schwinger bosons scales as $S^{-1}$. In contrast,  Eq.~(\ref{eq:green_cond}) indicates that Green's function of the condensed Schwinger bosons, $\mathcal{G}_c$  scales as $S^{0}$  (note that $\Omega_c = {h\over 2} \sim S $). This analysis provides  a $1/S$ power counting rule for evaluating relative contributions of different Feynman diagrams at the SP level [see Fig.~\ref{powercounting}~(a), (b), (f)].

The classical limit is then dominated by the contributions from the condensed spinons. For instance,  the SP contribution to the ground state energy per site becomes\cite{Ceccatto1993}
\begin{equation}
E_{sp}=\sum_{{\bm \delta}>0} J_{\delta} \left( B^2_{\bm \delta} -|A_{\bm \delta}|^2 \right) \to -{3\over 2}JS^2, 
\end{equation}
which corresponds to the classical limit ($S\to \infty$).
The magnetic moment becomes $n_c/2=S$, which is also the expected value in the classical limit.

\begin{figure}[t!]
\centering
\includegraphics[scale=0.42]{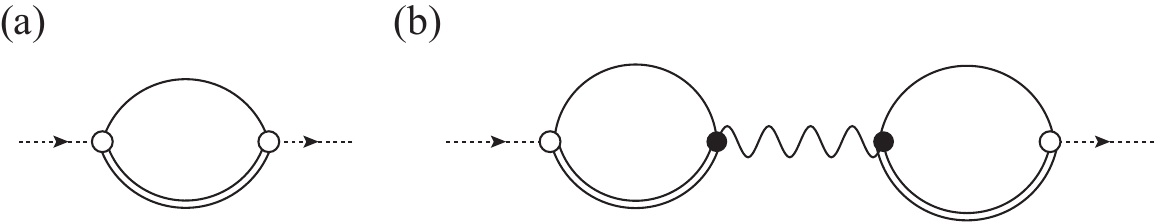}\caption{ Feynman diagrams of the dynamical structure factor in the large $S$ limit. (a) Saddle point contribution. (b) $1/N$ diagram that account for the true collective modes (magnons) of the magnetically ordered ground state. These modes appear as poles of the RPA propagator represented as a wavy line.}
\label{feynman}
\end{figure}

\subsubsection{Corrections beyond the Saddle point level}

 As $\mathcal{G}^{-1}$ is linear in the fields $\phi_{\alpha}$, the internal vertex turns out to be of order $S^0$:
 $v_{\alpha}=\frac{\partial \mathcal{G}^{-1}}{\partial \phi_{\alpha}}\sim S^0$.
 The RPA  propagator of the fluctuation fields can be expressed as $D^{-1}(\bm q,i\omega_n) = 2[ \Pi_0 - \Pi(\bm q, i\omega_n) ]$, where\cite{Ghioldi2018}

 \begin{equation}
 \Pi_{\phi_{\alpha_1} \phi_{\alpha_2}}(\bm q, i\omega_n) = \frac{1}{8} \tr\left[\mathcal{G}_{\rm sp} \; v_{\phi_{\alpha_1}} \! \mathcal{G}_{\rm sp} \ v_{\phi_{\alpha_2}} \right] \label{polarization}
 \end{equation}
 
\noindent is the polarization operator and $\Pi_0$ is a diagonal matrix containing the inverse of the exchange couplings $1/4J_{ij}$  along the diagonal except for the entries corresponding to $\lambda-\lambda$ derivatives, which are zero. By replacing the  Green's function (\ref{eq:GF_condensate}) in the polarization operator (\ref{polarization}) and  applying the power counting rule shown in Fig.~\ref{powercounting}, we obtain $\Pi_{\alpha\beta}(\bm q, i\omega_n) \sim S^0$ in the large $S$ limit (the dominant contribution arises from a loop containing  one condensed and one non-condensed spinon propagator).
It is then clear that $D(\bm q,i\omega_n)$ is ${\cal O}(S^0)$ in the large $S$ limit (Fig.~\ref{powercounting}(c)).

The resulting power counting rule for each Feynman diagram is  $(1/S)^{P_{nc}-L_{\Sigma}}$  where $P_{nc}$ is the number of propagators of {\it non-condensed} bosons and $L_{\Sigma}$ is the number of independent loops (i.e.,  the number of independent integrals over frequency).

\begin{figure}[t!]
\centering
\includegraphics[scale=1.1]{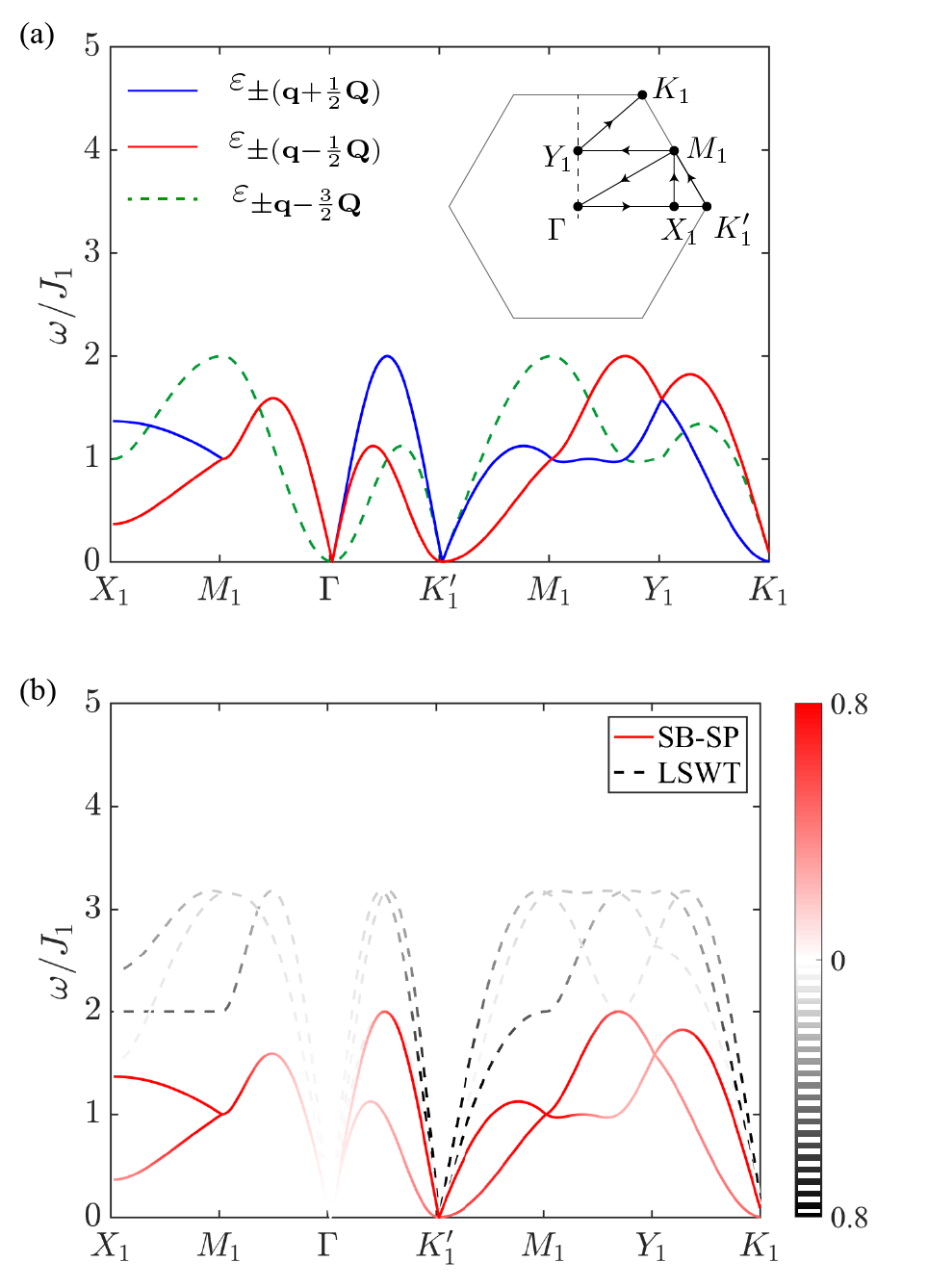}\caption{ (a) Dispersion relation of the poles of the dynamical spin susceptibility in the laboratory reference frame at the SP level. Each line is doubly-degenerate.
The spectral weight (residue of the pole) is zero for the dashed lines that correspond to in-plane modes [see panel (b)], while it is finite for the full lines that correspond to in-plane and out of plane modes. (b) Dynamical spin structure factor obtained from the SBMFT (red line) and from LSWT (black line). The color scale represents the spectral weight.}
\label{Ssp}
\end{figure}

\section{Dynamical spin structure factor}
\label{DSS_largeS}

We  are now ready to take the large-$S$ limit of the $T=0$ dynamical structure factor for the physical $SU(2)$ ($N=2$) version of the  spin model:
\begin{equation}
{S}_{\mu \nu} ({\bm q}, \omega) = - \frac{1}{\pi} \Im [\chi_{\mu \nu} ({\bm q}, \omega)],
\end{equation}
The off-diagonal components vanish for symmetry reasons.
At the SP level, the magnetic susceptibility is obtained by an analytic continuation $i\omega_n \rightarrow \omega + i0^+$ of ${\chi}^{sp}_{_I \mu\nu}(\bm q, i\omega_n)$ given in
Eq.~\eqref{chispI}, which corresponds to the diagram shown in Fig.~\ref{fig2:feymann}~(a). Along the $\omega$-axis, the imaginary part of ${\chi}^{sp}_{_I \mu\nu}(\bm q, \omega)$ includes a two-spinon continuum arising from two non-condensed spinons [spinon lines in Fig.~\ref{feynman}~(a) with momentum $\bm k + \bm q$ and ${\bm k}$ are both non-condensed bosons] and $\delta$-peaks
arising from one condensed spinon with ${\bm k} = {\bm 0}$ and one non-condensed spinon with momentum ${\bm k}=\pm {\bm q}$ in Fig.~\ref{feynman}~(b). The resulting dispersion of these $\delta$-peaks is $\varepsilon_{\pm {\bm q} -{\bm Q \over 2}}$.  The in-plane components of the dynamical structure factor, $S^{xx}({\bm q},\omega)$ and $S^{yy}({\bm q},\omega)$, contain four $\delta$-peaks centered at $\varepsilon_{\pm {\bm q} + {\bm Q \over 2}}$ and $\varepsilon_{\pm {\bm q} - {3 \bm Q \over 2}}$ for each ${\bm q}$, while  the out-of-plane component, $S^{zz}({\bm q},\omega)$, contains two $\delta$-peaks centered at $\varepsilon_{\pm {\bm q} -{\bm Q \over 2}}$ for each ${\bm q}$. Due to inversion symmetry, the six $\delta$-peaks form three groups of degenerate pairs [see Fig.~\ref{Ssp}~(a)].

The weight of the two-spinon continuum vanishes in the large-$S$ limit because $\mathcal{G}^{sp}_0 \sim S^{-1}$ and $\mathcal{G}^{sp}_{c} \sim S^{0}$. The remaining $\delta$-peak contributions (corresponding to the poles of the SBMFT) lead to a single-particle spectrum, which is {\it qualitatively different from the single-magnon spectrum of the LSWT} (see Fig.~\ref{Ssp}). The first qualitative difference appears in the number of gapless modes. While the single-magnon spectrum of the LSWT has only three gapless (Goldstone) modes, which are linear in momentum, the dispersion relation of the  poles obtained at the SP level also includes  spurious  {\it quadratic} modes that become gapless in the $S\to \infty$ limit. The second qualitative  difference  appears in the velocities of the three linear Goldstone modes. The LSWT correctly predicts that the velocity of the Goldstone mode associated with spin rotations along the axis perpendicular  to the plane defined by the coplanar spin ordering is different from the velocity of the other two Goldstone modes associated with the two independent in-plane rotations. In contrast, the three linear modes of the magnetic excitations that are obtained at the SP level have exactly the same velocity. This unexpected degeneracy  arises from a simple fact: the four gapless spinon modes have  the same velocities because of the  invariance of  ${\cal H}$ under global SU(2) spin rotations (for the moment we are focusing on $N=2$), that connect the spin up with the spin down spinon branches, and under spatial inversion that connects the spinon branches $\pm {\bm Q}/2 $ in the original reference frame, while preserving the spin quantum number. This degeneracy can be associated with an enlarged $O(4)$ symmetry that emerges upon taking the long wavelength limit of the SP action. This O(4) symmetry is broken by the $1/N$ corrections to the action, i.e., by the terms that account for the  interactions between spinons mediated by fluctuations of the gauge fields.
Note that these interaction terms become  irrelevant at the quantum critical point that signals the transition into a spin liquid state, implying that the fixed point that describes this transition has an emergent  O(4) symmetry.~\cite{Chubukov1994} For arbitrary values of $N$, the emergent symmetry becomes $O(2N)$.

In summary, going back to $N=2$, the SP result provides an incorrect description of the  low-energy modes of non-collinear ordered magnets because: i) the low-energy spectrum includes extra unphysical or spurious gapless modes with a quadratic dispersion and  ii) the three Goldstone modes form a degenerate triplet due to an extra ``isospin'' SU(2) symmetry, which is broken by the higher order terms in the $1/N$ expansion. This symmetry analysis demonstrates that the correct collective modes (magnons) can only be obtained from the SB theory  by going beyond the SP level.~\cite{Ghioldi2018}

To see how the above-mentioned analysis is reflected by our calculations, we note that after taking  the $S\to \infty $ limit,   $\varepsilon_{\bm q}$ includes two gapless modes at ${\bm q} \pm 3 {\bm Q}/2$ with a {\it quadratic dispersion}, in addition to the gapless modes with linear dispersion at ${\bm q} \pm  {\bm Q}/2$. The quadratic modes have a finite energy gap for finite $S$ values, while the linear modes remain gapless for arbitrary values of $S$. Given that  $\varepsilon_{\pm {\bm q} + {\bm Q \over 2}}$, $\varepsilon_{\pm {\bm q} - {3 \bm Q \over 2}}$ and $\varepsilon_{\pm {\bm q} -{\bm Q \over 2}}$ correspond to shifts of $\varepsilon_{\bm q}$ by three different wave-vectors, the $\delta$-peaks of the dynamical structure factor should also exhibit  linear and the quadratic gapless modes. Indeed, as indicated in
Fig.~\ref{Ssp}~(a), the gapless modes appear at the $\Gamma$ point and at the $K$ points (ordering wave vector ${\bm Q}$) of the Brillouin zone.  
The two in-plane modes at $\varepsilon_{\pm {\bm q}-{3 \over 2} \bm Q}$, indicated with dashed lines in Fig.~\ref{Ssp}~(a), have no spectral weight. Consequently, as it is shown in Fig.~\ref{Ssp}~(b), the dynamical structure factor exhibits only two different doubly-degenerate gapless modes.  Both of them are linear at the $\Gamma$ point, while one is linear and the other one is quadratic  at the $K_1$ and $K_1^{\prime}$ points. In contrast,  Fig.~\ref{Ssp}~(b) shows the three  linear Goldstone modes at the $\Gamma$, $K_1$ and $K_1^{\prime}$ points that appear in the dynamical structure factor of the  LSWT. As expected from the  $O(4)$ symmetry of the SP action, the linear spinon modes have the same velocity $v={3\over 2} J S$ at both the $\Gamma$ and $K$ points, while the Goldstone modes of the  LSWT have velocities $v_{\Gamma}={3\sqrt{3}\over 2} JS$ and $v_{K}={3\over 2} \sqrt{3\over2} JS$ at the $\Gamma$ and $K$ points, respectively. 
\begin{figure}[t!]
\centering
\includegraphics[scale=0.9]{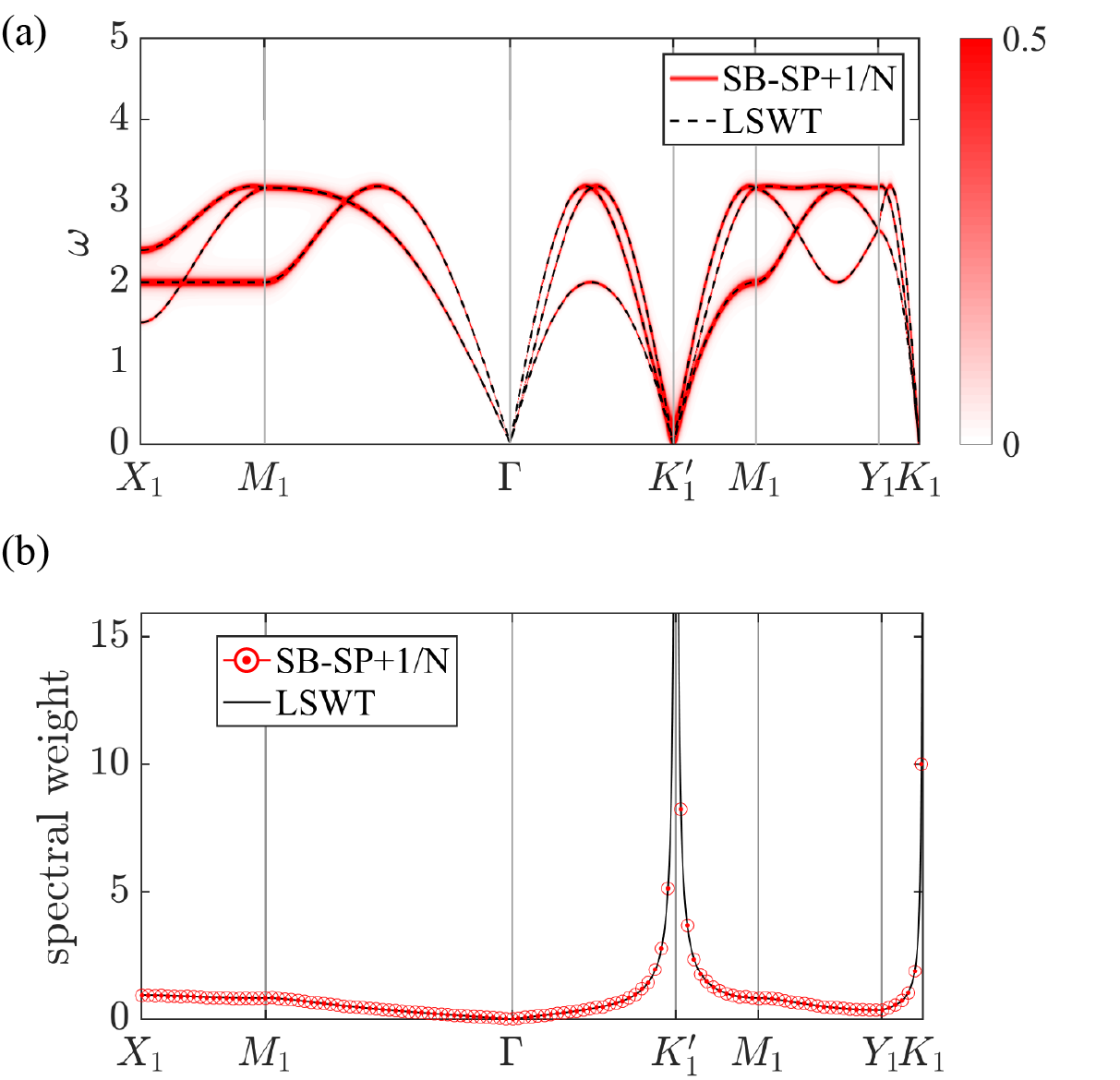}\caption{Dynamical spin structure factor obtained  from the Schwinger boson theory by including the diagrams shown in Figs.~\ref{feynman}~(a) and (b) (red). The black lines correspond to the result from LSWT.
Panel (a) shows the magnon dispersion relation (poles of the dynamical spin structure factor), while panel (b) shows the momentum dependence of the intensity of the magnon peak.}
\label{Sgs}
\end{figure}

The key observation of this work is that the correct dynamical spin structure factor in the large-$S$ limit is recovered only after adding  the $1/N$ correction corresponding to the diagram shown in Fig.~\ref{feynman}(b). Note that both diagrams in Figs.~\ref{feynman}~(a) and (b) are of order $S^0$. The effect of this $1/N$ correction is twofold: it cancels out exactly the poles of the SP contribution (the quadratic and the linear ones), while a new quasi-particle peak (delta function) emerges from the  pole of the RPA propagator of the fluctuation fields [note that the poles of the RPA propagator are also poles of the diagram shown in Fig.~\ref{fig2:feymann}~(b).~\cite{Ghioldi2018} 
 The cancellation of the SP contribution along the spinon dispersion, i.e., on the shell  $\omega= \omega_{\pm}=\varepsilon_{\pm {\bm q}-\frac{\bm Q}{2}}$,  for the $zz$-component of the dynamical spin susceptibility  can be  derived as follows.~\cite{Raykin1993,Auerbach1994} 
After noticing that the trace in Eq.~\eqref{corr}  reduces to $ \tr\big[ \mathcal{G}^{\rm sp} \ v_{\phi_{\alpha_1}} \ \mathcal{G}^{\rm sp} \ u^{z} \big]=\tr\big[ \mathcal{G}^{\rm sp}_0 \ v_{\phi_{\alpha_1}} \ \mathcal{G}^{\rm sp}_c \ u^{z}  \big] + \tr\big[ \mathcal{G}^{\rm sp}_c \ v_{\phi_{\alpha_1}} \ \mathcal{G}^{\rm sp}_0 \ u^{z}  \big] $,~\cite{Ghioldi2018} it is possible to demonstrate that 
 \begin{eqnarray}
&& \lim_{\omega \to \omega_{\pm}} (\omega- \omega_{\pm}) \tr\big[ \mathcal{G}^{\rm sp}_0 \ v_{\phi_{\alpha_1}} \ \mathcal{G}^{\rm sp}_c \ u^{z}({\bm q}, i \omega) \big]
\nonumber \\
&& =  \lim_{\omega \to \omega_{\pm}} (\omega- \omega_{\pm}) C_{\bm q} \left( \Pi_{\phi_{\alpha_1}, W_A }^{(c)} +\Pi_{\phi_{\alpha_1},\overline{W}_A}^{(c)} \right), \label{trace}
\end{eqnarray}
where $\Pi_{\phi_{\alpha_1},\phi_{\alpha_2}}^{(c)}= \frac{1}{8} \tr\big[ \mathcal{G}^{\rm sp}_0 \ v_{\phi_{\alpha_1}} \ \mathcal{G}^{\rm sp}_c \  v_{\phi_{\alpha_2}}  \big] + \frac{1}{8} \tr \big[ \mathcal{G}^{\rm sp}_c \ v_{\phi_{\alpha_1}} \ \mathcal{G}^{\rm sp}_0 \ v_{\phi_{\alpha_2}}  \big]$ and $C_{\bm q}$ is a ${\bm q}$-dependent proportionality constant.
Furthermore,  the RPA propagator can be safely approximated by $D(\bm q,i \omega_{\pm}) \approx [2 (\Pi_0 -\Pi^{(c)}) ]^{-1}({\bm q},i \omega_{\pm})$ in the large-$S$ limit.  Then, by replacing $D$ and the above trace in Eq. (\ref{corr}),  we find that  the poles of ${\chi}^{sp}_{_I zz}$ along the SP spinon dispersion are exactly cancelled with poles of  ${\chi}^{fl}_{_{II} zz}$  at the same frequency. A similar analysis can be applied to the in-plane, $xx$ and $yy$, components of the dynamical spin susceptibility. It is important to remark that this cancellation  holds for any value of $S$.~\cite{Ghioldi2018}\\

On the other hand, the poles of the RPA propagator are zeros of the fluctuation matrix (\ref{s2}):
\begin{equation}
S^{(2)}(\bm q, \omega) \cdot X =0.
\end{equation}
The first four components of  $X=(X_1(\bm \delta),X_2(\bm \delta),X_3(\bm \delta),X_4(\bm \delta),X_5)$ correspond to fluctuations of the Hubbard-Stratonovich fields $W^A_{\bm \delta}$, $\overline W^A_{\bm \delta}$, $W^B_{\bm \delta}$, $\overline W^B_{\bm \delta}$, respectively, and $X_5$ is the fluctuation of the Lagrange multiplier.

The pole equation turns out to depend on four linear combinations of $X$, namely
\begin{align}
R_{1}\equiv & \;c_{1}+c_{2}+c_{3}^{-}+c_{4}^{-},\\
R_{2}\equiv & \;c_{1}-c_{2}-c_{3}^{+}+c_{4}^{+}+4iX_{5},\\
R_{3}\equiv & \; \bar{c}_{1}+\bar{c}_{2}-\bar{c}_{3}^{-}-\bar{c}_{4}^{-},\\
R_{4}\equiv & \; \bar{c}_{1}-\bar{c}_{2}+\bar{c}_{3}^{+}-\bar{c}_{4}^{+}-4iX_{5},
\end{align}

\noindent where

\begin{align}
c_{1}(\bm{q})\!=\!\sum_{\bm{\delta}}\!F_{q}^{-*}(\bm{\delta})X_{1}(\bm{\delta}), & \;\; \;c_{2}(\bm{q})\!=\!\sum_{\bm{\delta}}\!F_{q}^{-*}(\bm{\delta})X_{2}(\bm{\delta}), \nonumber \\
c_{3}^{-}(\bm{q})\!=\!\sum_{\bm{\delta}}\!F_{q}^{-*}(\bm{\delta})X_{3}(\bm{\delta}), & \;\; \;c_{3}^{+}(\bm{q})\!=\!\sum_{\bm{\delta}}\!F_{q}^{+*}(\bm{\delta})X_{3}(\bm{\delta}),\nonumber \\
c_{4}^{-}(\bm{q})\!=\!\sum_{\bm{\delta}}\!F_{q}^{-*}(\bm{\delta})X_{4}(\bm{\delta}), & \;\; \;c_{4}^{+}(\bm{q})\!=\!\sum_{\bm{\delta}}\!F_{q}^{+*}(\bm{\delta})X_{4}(\bm{\delta}),\nonumber 
\end{align}
and
\begin{align}
\bar{c}_{1}(\bm{q})\!=\!\sum_{\bm{\delta}}\!\bar{F}_{q}^{-*}(\bm{\delta})X_{1}(\bm{\delta}), &\;\; \; c_{2}(\bm{q})\!=\!\sum_{\bm{\delta}}\!\bar{F}_{q}^{-*}(\bm{\delta})X_{2}(\bm{\delta}),\nonumber \\
\bar{c}_{3}^{-}(\bm{q})\!=\!\sum_{\bm{\delta}}\!\bar{F}_{q}^{-*}(\bm{\delta})X_{3}(\bm{\delta}), & \;\; \;c_{3}^{+}(\bm{q})\!=\!\sum_{\bm{\delta}}\!\bar{F}_{q}^{+*}(\bm{\delta})X_{3}(\bm{\delta}),\nonumber \\
\bar{c}_{4}^{-}(\bm{q})\!=\!\sum_{\bm{\delta}}\!\bar{F}_{q}^{-*}(\bm{\delta})X_{4}(\bm{\delta}), &\;\; \; c_{4}^{+}(\bm{q})\!=\!\sum_{\bm{\delta}}\!\bar{F}_{q}^{+*}(\bm{\delta})X_{4}(\bm{\delta}). \nonumber 
\end{align}
Here we have introduced the following two functions
\begin{align}
F_{q}^{\mp}(\bm{\delta}) & =e^{i(\bm{q}-\frac{\bm{Q}}{2})\cdot\bm{\delta}}\mp e^{i\frac{\bm{Q}}{2}\cdot\bm{\delta}},\nonumber \\
\bar{F}_{q}^{\mp}(\bm{\delta}) & =F_{-\bm{q}}^{\mp*}(\bm{\delta})=e^{i(\bm{q}+\frac{\bm{Q}}{2})\cdot\bm{\delta}}\mp e^{-i\frac{\bm{Q}}{2}\cdot\bm{\delta}}.\nonumber 
\end{align}

\noindent $R_1,...,R_4$ form a closed set of equations:

\begin{equation}
{\cal M}_{1} (\omega) \left(\begin{array}{c}
R_{2}\\
R_{4}
\end{array}\right)=3(1-\gamma_{\bm{q}})\omega
\left(\begin{array}{c}
R_{1}\\
R_{3}
\end{array}\right),\label{eq:le1}
\end{equation}

\begin{equation}
{\cal M}_{2} (\omega) \left(\begin{array}{c}
R_{1}\\
R_{3}
\end{array}\right)=\frac{3}{2}(1+2\gamma_{\bm{q}})\omega
\left(\begin{array}{c}
R_{2}\\
R_{4}
\end{array}\right).\label{eq:le2}
\end{equation}

\noindent where $\gamma_{\bm q} = {1\over 3} (\cos k_x + 2\cos {k_x \over 2} \cos {\sqrt{3} \over 2} k_y)$, and

\begin{equation}
{\cal M}_{1}\! (\omega)\!\!=\! 
\left( \!\! \begin{array}{cc}
\omega^2- \varepsilon_{\bm q -{\bm Q \over 2}}^2 - {1 \over 2} \omega_{\bm q}^2   & \omega^2- \varepsilon_{\bm q -{\bm Q \over 2}}^2  \\
\omega^2- \varepsilon_{\bm q +{\bm Q \over 2}}^2 & \omega^2- \varepsilon_{\bm q + {\bm Q \over 2}}^2 - {1 \over 2} \omega_{\bm q}^2
\end{array} \!\! \right),
\end{equation}

\begin{equation}
{\cal M}_{2}(\omega)=\left(\begin{array}{cc}
-\varepsilon_{\bm{q}-\frac{\bm{Q}}{2}}^{2} & \omega^{2}-\varepsilon_{\bm{q}-\frac{\bm{Q}}{2}}^{2}\\
\omega^{2}-\varepsilon_{\bm{q}+\frac{\bm{Q}}{2}}^{2} & -\varepsilon_{\bm{q}+\frac{\bm{Q}}{2}}^{2}
\end{array}\right).
\end{equation}

\noindent At $\omega\! =\! \omega_{\bm q}\!=3 \sqrt{(1-\gamma_{\bm q}) (1+2 \gamma_{\bm q})}$, the product of the two matrices is proportional to the two by two unit matrix

\begin{equation}
{\cal M}_{1}(\omega_{\bm q}){\cal M}_{2}(\omega_{\bm q})=\frac{1}{2}\omega_{\bm{q}}^{4}I_{2\times2}.
\end{equation}

\noindent Here we have used a simple relation between the single-spinon dispersion obtained from the SBMFT and $\omega_{\bm q}$:
\begin{equation}
\varepsilon^2_{\bm{q}-\frac{\bm{Q}}{2}} + \varepsilon^2_{\bm{q}+\frac{\bm{Q}}{2}}  =  {1\over 2} \omega_{\bm q}^2.\label{dispS} 
\end{equation}
In other words, Eqs.~(\ref{eq:le1}) and (\ref{eq:le2}) are satisfied for any choice of $R_{2}$, $R_{4}$ with $R_{1}$, $R_{3}$ determined by Eq.~(\ref{eq:le1}) when $\omega\!=\!\omega_{\bm q}$. Given that $\omega_{\bm q}$ is the single-magnon dispersion of the LSWT, 
this demonstrates that the  poles of the RPA propagator coincide with the poles of the LSWT [see Fig.~\ref{Sgs}~(a)].
In addition, as shown in Fig.~\ref{Sgs} (b), the spectral weight of  the magnon peak, defined as $W(\bm q)\! \!=\!\! \int d\omega S(\bm q, \omega)$, is also exactly captured by the two diagrams in Fig.~\ref{feynman}~(a) and (b). We note that there are other diagrams (or order $1/N$ and higher) that scale as $S^0$. Consequently, it is surprising that only the two diagrams in Figs.~\ref{feynman}~(a) and (b) are required to obtain the {\it exact} magnetic susceptibility in the large-$S$ limit.

\section{Discussion}
\label{Disc}

In summary, we have shown that it is necessary to go beyond the SP level of the Schwinger boson theory of the 
triangular lattice antiferromagnet in order to capture  the correct collective modes in the large-$S$ limit.
These modes  are two-spinon bound states generated by the interaction of spinons with the auxiliary fields (emergent gauge fields). The magnon energies are determined by the poles of the RPA propagator. This result must be contrasted with the dynamical susceptibility at the SP level, where the quasi-particle dispersion relation coincides with the single-spinon dispersion.

Although we have not shown it in this manuscript, this conclusion remains valid for the one-singlet bond AA decomposition \cite{Arovas1988,Read1991,Sachdev1991} of the Heisenberg interaction and for other non-collinear magnetic orderings  of frustrated Heisenberg Hamiltonians.
This result, along with the long wave-length limit of the $S=1/2$ theory that we presented in Ref.~[\onlinecite{Ghioldi2018}], demonstrate that the  Schwinger boson  theory can correctly capture the low-energy magnons of the underlying magnetically ordered state.
In addition, unlike the semiclassical $1/S$ expansion, the  Schwinger boson  theory  is well-suited for describing the higher energy continuum associated with the formation of two-spinon bound states (magnons) with  long confinement length scale. Given that this is the expected scenario for magnetically ordered states in the proximity of a QMP, we conclude that the Schwinger boson theory can be a more adequate tool for describing the spin dynamics of frustrated magnets with strong quantum fluctuations. 

While we have shown that the correct classical limit of the theory can be captured by including only the $1/N$ correction corresponding to the Feynman diagram of Fig.~\ref{fig2:feymann}~(b),  the  other $1/N$ diagrams of Fig.~\ref{fig2:feymann} may play a significant role in a {\it quantitative} description of the dynamical spin structure factor in the presence of strong quantum fluctuations. We note that the diagram shown in Fig.~\ref{fig2:feymann}~(c) corresponds to a vertex renormalization, while the  two diagrams shown in Figs.~\ref{fig2:feymann}~(d) and (e) correspond to a renormalization of the single-spinon propagator. 
In other words, we expect that these diagrams should renormalize the single-spinon dispersion along with the two-spinon continuum and the single-magnon (two-spinon bound state) dispersion. Magnon-magnon interaction effects are captured by diagrams of order $1/N^2$ and higher.~\cite{Ghioldi2018}

Finally, it is interesting to note that the situation is qualitatively different for {\it collinear} magnetic orderings of Heisenberg magnets, like the square lattice Heisenberg antiferromagnet, because of the residual U(1) symmetry group.
As it was explained in Ref.~\onlinecite{Ghioldi2018}, the bubbles of  the Feynman diagram shown in Fig.~\ref{fig2:feymann}~(b)
vanish for the transverse components of the dynamical susceptibility due to this U(1) symmetry. This cancellation implies that the $1/N$  contribution that we considered in this manuscript only corrects the longitudinal component of the magnetic susceptibility. In other words, unlike the case of the non-collinear orderings that we considered here, the SP contribution to the transverse components of the magnetic susceptibility is not corrected by the $1/N$ contribution shown in Fig.~\ref{fig2:feymann}~(b). However, it is still true that the poles of the RPA propagator coincide with the  single-magnon poles of the LSWT. We note  that the SP spinon dispersion is half of the single-magnon dispersion  in the large-$S$ limit: $\varepsilon_{\bm{q}+\frac{\bm{Q}}{2}}=\frac{1}{2} \omega_{\bm q}$. However, the missing factor of two is recovered, $\omega_{\bm q}=2\varepsilon_{\bm{q}+\frac{\bm{Q}}{2}}$, in the dispersion of the poles of the RPA propagator through equation \eqref{dispS} [$\varepsilon_{\bm{q}-\frac{\bm{Q}}{2}} = \varepsilon_{\bm{q}+\frac{\bm{Q}}{2}}$ for ${\bm Q}=(\pi,\pi)$)].\footnote{ Alternatively, an AA decomposition\cite{Arovas1988,Auerbach1994} of the Heisenberg term leads to a SP spinon dispersion that already coincides with the single-magnon dispersion: $\omega_{\bm q}=\varepsilon^{AA}_{\bm{q}+\frac{\bm{Q}}{2}}.$} It is also important to note that the SP expansions of collinear and non-collinear orderings {\it cannot} be continuously connected because the fluctuation matrix is not semi-positive defined around the Lifshitz transition point that connects both types of magnetic orderings.\cite{Manuel1999} In other words, the result that we presented here {\it cannot} be extended to collinear cases by taking the  collinear limit of a sequence of non-collinear magnetic orderings (continuous incommensurate to commensurate transition). Work to overcome the $U(1)$ residual symmetry problem for collinear antiferromagnets is in progress.

\begin{acknowledgments}  
We  thank C. J. Gazza for useful conversations. This work was partially supported by CONICET under grant Nro 364 (PIP2015).
S-S. Z. and C. D. B. were partially supported from the LANL Directed Research and Development program. Y. K. acknowledges the support from Ministry of Science and Technology (MOST) with the grant No. 2016YFA0300500 and No. 2016YFA0300501.
\end{acknowledgments}

%\bibliographystyle{apsrev4-1}
%\bibliography{ref}


\end{document}